\documentclass[aps,prb,superscriptaddress,twocolumn,floatfix]{revtex4}

\usepackage{amsmath}
\usepackage{amssymb}
\usepackage{bm,color}
\usepackage{graphicx,epsfig,color}
\usepackage{hhline}
\usepackage{ifthen}
\usepackage[utf8]{inputenc}

\usepackage{ulem}
\definecolor{orange}{RGB}{85,201,81}

\newcommand{\bra}[1]{\langle\,{#1}\, |}
\newcommand{\ket}[1]{|\,{#1}\,\rangle}
\newcommand{\braket}[2]{\mbox{$\langle\,{#1}\, | \,{#2}\,\rangle$}}

\newcommand{\lrb}[1]{\langle\, {#1}\,\rangle}

\newcommand{\adj}[1]{#1^{\dagger}}
\newcommand{\cc}[1]{{#1}^*}
\newcommand{\ii}{\mathrm{i}}

\newcommand{\ee}{\vec{e}}
\newcommand{\kk}{\vec{k}}

\newcommand{\ww}{\vec{w}}

\renewcommand{\H}[1]{H_\mathrm{#1}}

\newcommand{\kB}{k_{\rm B}}
\newcommand{\BathCor}{\alpha}

\newcommand{\Deriv}[2][\empty]{%
  \ifthenelse{\equal{#1}{\empty}}
    {D_#2}
    {D_{#1,#2}}
}

\newcommand{\psit}[1][\empty]{%
  \ifthenelse{\equal{#1}{\empty}}
    {\psi_t}
    {\psi_t^{(#1)}}
}
\newcommand{\npsit}[1][\empty]{%
  \ifthenelse{\equal{#1}{\empty}}
    {\tilde\psi_t}
    {\tilde\psi_t^{(#1)}}
}
\newcommand{\rhot}[1][\empty]{%
  \ifthenelse{\equal{#1}{\empty}}
    {\rho_t}
    {\rho_t^{(#1)}}
}

\newcommand{\TT}{TT}
\newcommand{\nPA}{$n$PA}
\newcommand{\nMA}{$n$MA}

\renewcommand{\emph}{\textit} 

\begin{document}

\title{Flexible scheme to truncate the hierarchy of pure states}

\author{P.-P. Zhang}
\affiliation{Max Planck Institute for the Physics of Complex Systems,
N\"othnitzer Strasse 38, D-01187 Dresden, Germany}

\author{C. D. B. Bentley}
\affiliation{Max Planck Institute for the Physics of Complex Systems,
N\"othnitzer Strasse 38, D-01187 Dresden, Germany}

\author{A.\ Eisfeld}
\affiliation{Max Planck Institute for the Physics of Complex Systems,
N\"othnitzer Strasse 38, D-01187 Dresden, Germany}

\date{\today}

\begin{abstract}
The hierarchy of pure states (HOPS) is a wavefunction-based method which can be used for numerically modeling open quantum systems.
Formally, HOPS recovers the exact system dynamics for an infinite depth of the hierarchy.
However, truncation of the hierarchy is required to numerically implement HOPS.
We want to choose a 'good' truncation method, where by 'good' we mean that it is numerically feasible to check convergence of the results.
For the truncation approximation used in previous applications of HOPS, convergence checks are numerically challenging.
In this work we demonstrate the application of the '$n$-particle approximation' ($n$PA) to HOPS.
We also introduce a new approximation, which we call the '$n$-mode approximation' ($n$MA).
We then explore the convergence of these truncation approximations with respect to the number of equations required in the hierarchy.
We show that truncation approximations can be used in combination to achieve convergence in two exemplary problems: absorption and energy transfer of molecular aggregates. 
\end{abstract}
\maketitle

\section{Introduction}

Open quantum system approaches have become increasingly popular in the description of large assemblies of coupled  molecules, which are interacting with their surroundings (like the solvent or a protein); for some examples see Refs.~ \cite{KReMa96_99_,IsFl09_234111_,MoReLl08_174106_,PlHu08_113019_,RoEiWo09_058301_}.
Often it is possible to choose as the system part only electronic states of the molecular assembly; molecular vibrational modes and the effect of the surroundings are then modeled as an environment of harmonic oscillators, linearly coupled to system states (see e.g.~\cite{VaEiAs12_224103_,RoStWh12_204110_,ChAgVa15_9995_}).
This environment then typically exhibits so-called non-Markovian behaviour.
It is difficult to treat this problem numerically.
The application of popular approaches, based on Lindblad or Redfield \cite{MaKue11__} equations,
is quite limited.
For example, they cannot capture the effect of strong coupling to weakly damped vibrational modes. 

One method to handle this problem, that is used extensively, is the so-called hierarchical equation of motion (HEOM) approach \cite{Ta90_6676_,Ta06_082001_,IsFl09_17255_,StSc09_225101_,KrKrRo11_2166_}.
In this method the reduced density matrix of the system is obtained by solving a coupled system (hierarchy) of differential equations of density-matrix-like objects.
One drawback of this approach is that  the required number of differential equations rapidly grows with the number of (weakly damped) vibrational modes and upon decreasing the temperature. 
Another drawback is that the size of the density matrix grows quadratically with the system size.

To overcome the problems associated with the HEOM approach, in recent years numerically efficient approaches to calculate the reduced density matrix using stochastic wavefunctions within the non-Markovian Quantum State Diffusion (NMQSD) \cite{DiSt97_569_,DiGiSt98_1699_,YuDiGi99_91_} framework have been developed \cite{RoEiWo09_058301_,SuEiSt14_150403_,LiYiDe14_022122_,HaSt17_5834_,Ritschel}.
In particular in Ref.~\onlinecite{SuEiSt14_150403_} a stochastic hierarchy of pure states (HOPS) was developed, with which one can recover the reduced density matrix exactly (for an infinite number of trajectories and infinite depth).
For the case of excitation transfer in light harvesting systems it was demonstrated that for typical parameters one has fast convergence with respect to the number of trajectories and the depth of the hierarchy.
The HOPS is closely related to HEOM.
In Ref.~\onlinecite{SuStEi15_1408_} it was shown that HEOM can be directly derived from HOPS.
Our studies indicate that as well as reducing the size (wavefunction versus density matrix), HOPS also converges faster with the depth of the hierarchy (this is related to the fact that the $n$-th order of HOPS already contains terms that appear in the $n^2$th order of HEOM).

As in HEOM, the number of equations appearing in HOPS grows with the depth of the hierarchy and the number of 'modes'.
Therefore, one would like to have a flexible scheme to truncate the hierarchy in such a way that one still has a numerically treatable problem while checking for convergence.
We desire in particular that one can increase the size of the numerical system of equations gradually in steps that are not too large (the steps should also not be too small because of computational overhead).
One way of efficiently truncating the hierarchy is based on the so-called n-particle approximation (\nPA{}) \cite{npa, Ph71_2039_}, which was recently also adapted to HEOM \cite{SoBaSh15_064109_}.
The two-particle approximation (2PA, or TPA) has for example been used extensively to treat molecular aggregates like self-assembled organic dyes in solution \cite{TPA_MCTDH}, molecular crystals\cite{KlJo80_25_}, two-dimensional monolayers \cite{Sp04_7643_}, carotenoid assemblies \cite{Sp09_4267_}, and photosynthetic light harvesting systems \cite{TPA_LHC_2,TPA_LHC}.

In the present work we first show that the \nPA{ }also works for HOPS, as expected.
Then we present a new, even more flexible scheme, which we denote by n-mode approximation \nMA.
Using a combination of \nPA{ }and \nMA{ }allows for sufficient flexibility in performing convergence checks.

The paper is organized as follows: in section~\ref{sec:Method}, we review the HOPS method and describe the molecular open quantum system used in this work.
In section~\ref{sec:trunc}, we present different truncation approximations for the hierarchy.
We then assess the quality of the different approximations in section~\ref{sec:quality}.  We do this by applying the truncation approximations with different order to calculate absorption spectra and energy transfer in the molecular aggregate (open quantum system).  
This allows us to discuss convergence of the truncation schemes with an increasing number of equations in the hierarchy.
We conclude in section~\ref{sec:conc}.

\section{Method}
\label{sec:Method}
We first review the HOPS approach to open quantum system dynamics and then describe how we use it to calculate 2D spectra.
\subsection{Open quantum system model}

We consider the (total) Hamiltonian
\begin{equation}
  \H{tot} = H  + \H{env} + \H{int},
  \label{eq:H_tot}
\end{equation}
where $H$ is the Hamiltonian of the 'system',
which for the molecular aggregate (with $N$ molecules) considered here reads:
\begin{eqnarray}\label{eq:H_sys}
\begin{aligned}
H=\sum_{\ell=1}^{N}\varepsilon_{\ell} |\ell\rangle\langle \ell |+\sum_{\ell,\ell'=1}^{N}V_{\ell \ell'}|\ell \rangle\langle \ell'|.
\end{aligned}
\end{eqnarray}
It contains the electronic excitations of the molecules and their mutual interactions.
For simplicity we consider only states with a single molecule excited, which is sufficient to describe electronic excitation transfer and linear optical spectra.
In Eq.~(\ref{eq:H_sys}) the states $\ket{\ell}$ denote states where molecule $\ell$ is electronically excited and all the others are in the electronic ground state (we take two electronic states per molecule into account).
The transition energies of molecule $\ell$ are denoted by  $\varepsilon_{\ell}$ and the transition dipole-dipole interaction is $V_{\ell m}$.

The Hamiltonian of the environment is given by
\begin{equation}
  \H{env} =\sum_{\ell} \H{env}^{(\ell)}= \sum_{\ell}\sum_{ \lambda} \omega_{\ell \lambda} \adj{b}_{\ell \lambda} b_{\ell \lambda}
\end{equation}
 consisting of harmonic oscillators (
 $ \big[ b_{\ell \lambda}, b_{\ell \lambda'} \big] = 0$
and
  $\big[ b_{\ell \lambda}, \adj{b}_{\ell' \lambda'} \big] = \delta_{\ell \ell'} \delta_{\lambda\lambda'}$).
Here we have partitioned the environment into independent parts for each molecule labeled by the index $\ell$.
The interaction of system and environment is modeled by a linear coupling Hamiltonian
\[
  \H{int} = \sum_{\ell} \sum_{\lambda} \left( \cc{g}_{\ell \lambda} L_{\ell} \adj{b}_{\ell \lambda} + g_{\ell\lambda} \adj{L_{\ell}} b_{\ell\lambda} \right).
\]
Here, $L_{\ell}\equiv \ket{\ell}\bra{\ell}$ is a system operator that couples to the $\ell$th environment and $g_{\ell \lambda}$ are complex numbers quantifying the coupling strength of the respective oscillator  $(\ell,\lambda)$ to the system.

It is convenient to encode the frequency dependence of the interaction strengths in the so-called spectral densities
\[
  C_{\ell}(\omega) = \sum_{\lambda} |g_{\ell \lambda}|^2 \delta(\omega-\omega_{\ell \lambda}),
  \label{eq:spectral_density}
\]
which are typically assumed to be continuous functions of frequency.
The latter is related to the bath correlation function $\BathCor_{\ell}(\tau)$ by \cite{MaKue11__}
\begin{equation}
  \label{eq:BathCor}
  \BathCor_{\ell}(\tau)=\! \!  \int_0^{\infty}\! \!d\omega\, C_{\ell}(\omega)\Big(\coth \! \! \big(\frac{ \omega}{2 \kB T}\big)\, \cos(\omega \tau)
   - i \sin(\omega \tau) \Big)
\end{equation}
where $T$ is the temperature. 
In many cases of interest, the bath-correlation function can be well approximated by a sum of exponentials \cite{MeTa99_3365_,DaWiPo12__,RiEi14_094101_}: 
\begin{equation}
\begin{split}
\label{eq:bathcorExp}
\BathCor_{\ell}(\tau)=& \sum_{j=1}^J p_{\ell j} \exp(-w_{\ell j} t) \quad \quad ;(t>0)
\end{split}
\end{equation}
with $w_{\ell j}=i \Omega_{\ell j} +\gamma_{\ell j}$. 
Here we denote the number of exponentials in the sum by $J$.
Such a decomposition allows the derivation of a hierarchy of coupled equations.

\subsection{The HOPS approach}
The basic equation of the HOPS approach is a stochastic hierarchy of differential equations \cite{SuEiSt14_150403_,SuStEi15_1408_}, which for the molecular aggregate described in the previous section takes the form  (here and in the following we use $\psi_t=\psi(t)$ interchangeably)
\begin{eqnarray}
  \partial_t \psit[\kk](\mathbf{z})
  &=& \left( -\ii H - \kk\cdot\ww + \sum_{\ell j} z^*_{\ell j}(t) L_{\ell} \ \right) \psit[\kk](\mathbf{z}) \nonumber\\
  &&+ \sum_{\ell j}  k_{\ell j} p_{\ell j} L_{\ell } \psit[\kk - \ee_{\ell j}](\mathbf{z})\label{eq:hops}\\
&&
  - \sum_{\ell j}  \adj{L}_{\ell} \psit[\kk + \ee_{\ell j}](\mathbf{z}) \nonumber
\end{eqnarray}
with initial conditions $\psi^{(\vec{0})}_{t=0}=\psi_{t=0}$ and  $\psi^{(\kk)}_{t=0}=0$ for $\vec{k}\ne 0$.  
The $\mathbf{z}=\mathbf{z}_t$ are a set of complex stochastic processes with $\mathcal{M}_z \{z\}=0$ and $\mathcal{M}_z \{z_{\ell j}(t) z^*_{\ell j}(s)\}=\BathCor_{\ell j}(t-s)$.
Here  $\mathcal{M}_z $ denotes an average over the stochastic wavefunctions,
$\ww=\{w_{1,1},\dots,w_{N,J}\}$, where $J$ denotes the number of exponentials in Eq.~(\ref{eq:bathcorExp}), and 
\begin{equation}
\label{eq:kk}
\kk=\{k_{1,1},\dots,k_{N,J}\} 
\end{equation}
 with   $k_{\ell j}$ integers $\ge 0$.
Furthermore, $\ee_{\ell j}=\{0,\dots,1,\dots 0\}$ is a vector that has a one at the $(\ell,j)$th position and the rest of the elements are zero.
 The numbers $k_{\ell j}$ can be interpreted as the number of excitations of the respective mode $(\ell,j)$ of decomposition of the bath correlation function Eq.~(\ref{eq:bathcorExp}). 

Equation (\ref{eq:hops}) is strictly valid for a bath-correlation of the type of Eq.~(\ref{eq:bathcorExp}). 
In practice we do not directly use Eq.~(\ref{eq:hops}) but use the corresponding non-linear equation that has much better convergence properties with respect to the number of trajectories (see the discussion in Ref.~\cite{SuEiSt14_150403_})

 Expectation values of an operator $A$ in the system space can be obtained via
\begin{equation}
\label{eq:expectation}
\lrb{A}=\mathcal{M}_z \{\bra{\psi(t;\mathbf{z})} A\ket{\psi(t;\mathbf{z})}\}
\end{equation}
The quantity entering the expectation value Eq.~(\ref{eq:expectation}) is $\psi(t)=\psi^{(\vec{0})}(t)$.

When considering excitation transfer, we are in particular interested in the time dependent probabilities to find excitation on a certain molecule.
The respective operators are the projectors $L_{\ell}=\ket{\ell}\bra{\ell}$.

\subsection{Absorption}
For linear optical properties like absorption, or linear and circular dichroism, it turns out that one can use the same hierarchy Eq.~(\ref{eq:hops}), but it is sufficient to consider only a single trajectory where all $z_{\ell j}(t)\equiv 0$.
Details can be found in Ref.~\onlinecite{RiSuMoe15_034115_}.

\section{Truncation} \label{sec:trunc}
The hierarchy Eq.~(\ref{eq:hops}) consists of an infinite number of coupled equations.
For numerical implementations one has to truncate the hierarchy (and also has to  consider only a finite number of stochastic trajectories).   
One is then interested in obtaining results within a certain accuracy.

For an efficient implementation one wants to keep the number of coupled equations as small as possible for the desired accuracy.
Here it is essential to have a 'good' truncation procedure.
In the following we will illustrate this point by considering three different truncation schemes, which we will denote by '{\it triangular truncation}' (\TT), the '{\it n-particle approximation}' (\nPA) and the '{\it n-mode approximation}' (\nMA).

One has to keep in mind that it is {\it a priori} not clear how good a specific approximation is.
A large number of auxiliary states (equations) does not necessarily mean a better accuracy of the result, since the auxiliary states may not contain the relevant ones.
Therefore, we will also consider the quality of the different approximations.

\subsection{Triangular truncation (\TT{})}
The \TT{} is a  simple truncation scheme.
Here one takes all terms of Eq.~(\ref{eq:hops}) into account that fulfill the condition 
\begin{equation}
\label{eq:TT}
\sum_{\ell=1,j=1}^{NJ}k_{\ell j}\le D
\end{equation}
where $D$ is a positive integer.
In the case of the equality, the last term on the right hand side of Eq.~(\ref{eq:hops}) is then suitably approximated, using only lower orders \cite{SuEiSt14_150403_}. 
In the present work we simply set this so-called terminator to be zero, i.e., we ignore the final line in Eq.~(\ref{eq:hops}).
Previous works with HOPS have always used this TT scheme (with a slightly more sophisticated terminator) \cite{SuEiSt14_150403_,SuStEi15_1408_,RiSuMoe15_034115_}. 
Convergence is checked by increasing $D$ and recording the difference between the results for $D$ and $D-1$.
Details on such convergence checks can be found in the supporting material of Ref.~\onlinecite{SuEiSt14_150403_}.

A drawback of this scheme is that for a large number $N$ of molecules and a large number of modes $J$, the number of equations increases very fast:
\begin{eqnarray}\label{eq:TT_numbers}
\begin{aligned}
M_{\mathrm{\TT}}= \sum_{d=1}^{D}\dbinom{d+NJ-1}{d} 
\end{aligned}
\end{eqnarray}
For example for $N=10$ and $J=5$ one has for $D=1$ a moderate number of equations $M_{\mathrm{\TT}}= 46$, but already for $D=2$ one has quite a large number $M_{\mathrm{\TT}}= 1127$. 
This can also be seen in Fig. \ref{D_vs_M}.

In the same spirit as Eq.~(\ref{eq:TT}) one can also use truncation conditions that take specifics of the modes into account.
For example one expects that for weakly-coupled modes (small $p_{\ell j}$) or strongly-damped modes (large $\gamma_{\ell j}$) one does not need a large 'excitation' and one could use 
\begin{equation}
\label{eq:TT2}
\sum_{\ell=1,j=1}^{NJ}\frac{\gamma_{\ell j}\Omega_{\ell j}}{|p_{\ell j}|}k_{\ell j}\le \tilde{D}.
\end{equation}
 We will not discuss this truncation scheme Eq.~(\ref{eq:TT2}) in the following. 
Our focus will be on Eq.~(\ref{eq:TT}), on which we impose further restrictions.

\subsection{$n$-particle approximation (\nPA{})}
\label{sec:nPA}
The basic idea of the \nPA{} is that only terms with {\it at most} $n$ molecules that have vibrational excitation are taken into account.
As mentioned in the introduction a similar type of approximation has been extensively used and tested for linear molecular aggregates with one undamped vibrational mode per molecule \cite{Ph71_2039_,SpZhMe04_10594_,DiSpBo12_69_,Sp09_4267_,TPA_MCTDH,TPA_LHC,TPA_LHC_2,TPA_2D,TPA_2D_2}.
For the hierarchy Eq.~(\ref{eq:hops}) the \nPA{ }implies that tuples $\vec{k}$ are only taken into account when $k_{\ell j} \neq 0$ for no more than $n$ molecules. 
Let us write $\vec{k}= \{\vec{k}_1,\dots\vec{k}_N \}$ with $\vec{k}_\ell =\{k_{\ell 1},\dots, k_{\ell J}\}$.
For the one particle approximation (OPA=1PA), $n=1$, one then only takes into account the terms
$ \{\vec{k}_1,\vec{0}\dots,\vec{0} \}$, $ \{0, \vec{k}_2,\vec{0}\dots,\vec{0} \}$, \dots  $ \{0,\dots,\vec{0}, \vec{k}_N \}$.
Similarly, for the two particle approximation (TPA=2PA), $n=2$, one uses only terms of the form  $ \{\dots, 0,\ \vec{k}_{\ell'},\,\vec{0}\dots,\vec{0},\ \vec{k}_{\ell''},\,\vec{0}\dots \}$.

Clearly, one still needs to truncate the hierarchy. 
To do so one can still use the triangular conditions Eq.~(\ref{eq:TT}) or Eq.~(\ref{eq:TT2}). 
\begin{eqnarray}\label{eq:TPA numbers}
\begin{aligned}
M_{\mathrm{nPA }} =\sum_{d=1}^D \sum_{k=0}^{n-1}\Bigg[&\dbinom{N}{n-k}\dbinom{d+(n-k)J-1}{d}\\
&\times\dbinom{N-n-1+k}{k}(-1)^k
\Bigg]
\end{aligned}
\end{eqnarray}

Typically one reduces the number of equations even further, by requiring that one of the non-zero molecular $k$-vectors must belong to the molecule which is electronically excited. To make this clearer, note that if one writes Eq.~(\ref{eq:hops}) in the basis of localized excitation $\ket{\ell}$, then there will be terms like $\braket{\ell}{\psi^{\vec{k}}_t(\mathbf{z})}$.  
Now, conditioned on the index $\ell$, one of the corresponding non-zero molecular $k$-vectors must belong to molecule $\ell$.
For example, for the 1PA one has for $\braket{\ell}{\psi^{\vec{k}}_t(\mathbf{z})}$ only one allowed $k$-vector: $\vec{k}=\{\vec{0},\dots, \vec{k}_{\ell}, \dots, \vec{0} \}$. 
This further restriction reduces the number of equations by a factor $N$.

\subsection{$n$-mode approximation (\nMA{})}
\label{sec:nMA}
Here one only takes {\it in total} a maximum of $n$ indices $k_{\ell j}$ which are non-zero. 

Again, one needs to truncate the hierarchy. 
To do so one can still use the triangular conditions Eq.~(\ref{eq:TT}) or Eq.~(\ref{eq:TT2}). 
Then, in the case of a truncation scheme according to  Eq.~(\ref{eq:TT}),  the total number of equations with the \nMA{} is:
	\begin{eqnarray}
	\label{eq:nMA_numbers}
	M_{\rm nMA}=\sum_{i=1}^{n}\dbinom{NJ}{i}\sum_{d=1}^{D}\dbinom{d-1}{i-1}.
	\end{eqnarray}

One can further reduce the number of equations by combining the \nMA{ }with the two-particle approximation (or in general with the \nPA).

\subsection{Scalings of the different truncation schemes}
\begin{figure}[tpb]
\begin{center}
\includegraphics[width=8cm]{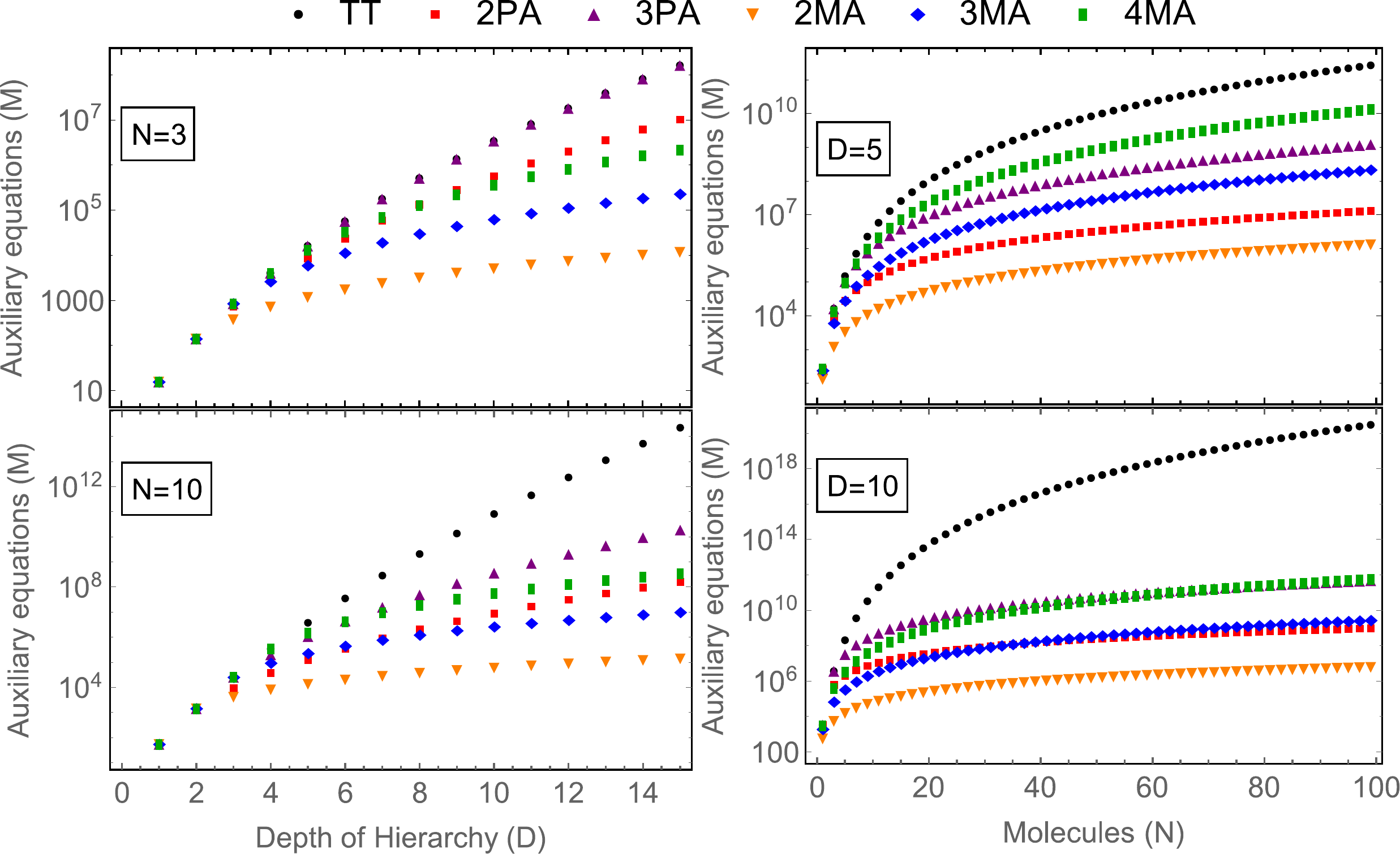}
\caption{\label{D_vs_M} Scaling of auxiliary equations $M$ with the depth of hierarchy $D$ (left column) and the number of monomers $N$ (right column).
In all cases $J=5$. The used value for $N$ (left column) or $D$ (right column) are provided in the plots.
The number of auxiliary equations is calculated according to the formulae Eq.~(\ref{eq:TT_numbers}), (\ref{eq:TPA numbers}) and (\ref{eq:nMA_numbers}).
Note the different ranges of the vertical axis.}
\end{center}
\end{figure}

To obtain a feeling for the number of equations one encounters for the different truncation schemes, in Fig.~\ref{D_vs_M} the scaling of the different truncation schemes with the hierarchy depth $D$, for fixed $J$ and $N$ values (left column) and the scaling with $N$ for fixed $J$ and $D$ are shown (right column).
In all cases we have used $J=5$.

Let us first take a look at the scaling with the depth $D$.
We show the two cases $N=3$ and $N=10$. 
One sees that for all cases the number of auxiliary equations $M$ grows quite rapidly.
At small depth there is little difference, however with increasing depth, the different approximations have different gradients of $M$ with $D$ (note that for the case $N=3$ the TT and the 3PA have identical sets of auxiliary equations).
One sees that the \nMA{} results in much smaller numbers than the respective \nPA{}.
Even for the case with small $N$ the various approximations result in order-of-magnitude differences in the number of auxiliary equations (note the logarithmic scale of the vertical axis).
For the case of larger $N$ this becomes even more extreme.

An important aspect is the scaling with the number of molecules $N$.
In the right column of Fig.~\ref{D_vs_M} we show examples for $D=5$ and $D=10$.
One sees an initial rapid growth with $N$, which then slows down.
For both cases shown the TT will be very difficult to handle numerically for more than approximately 20 molecules (for $D=5$ one then has $~10^8$ aux.~eq., and for $D=10$ even $10^{14}$). 
The schemes with a low number of excitations (in particular \nMA{} with $n<4$ and \nPA{} with $n<3$) still have reasonable numbers of auxiliary equations, even for very large aggregates with $N>100$ molecules.

\section{Quality of the different approximations} \label{sec:quality}

In the following we will consider some examples to investigate the quality of the different approximations.
We do this for two examples: 1.: Absorption spectra of a linear chain.
2.: Energy transfer in the photosynthetic Fenna-Matthews-Olson (FMO) complex.

We start with absorption, since there only a single trajectory is needed and the convergence with respect to the number of auxiliary equations is not complicated by the convergence with respect to the number of stochastic trajectories.

\subsection{Example 1: Calculation of absorption spectra}

\begin{table}[b]
    \centering
    \begin{tabular}{c*{2}{r}r}
        \hline
        $\quad j \quad $            &   $p_{j}\quad\quad$  &  $  \quad\Omega_j\quad $ &$\quad \gamma_j \quad$ \\
        \hline
        1         &     $24,000 - 660\,i$   &  $500$ &  5\\
        2         &     $275,000+660\,i$  &  $-500$ & 5 \\
        3         &    $-520$                   & 0        &1620
    \end{tabular}
    \caption{\label{tab:BCF_abs_params} The parameters used in the bath correlation function Eq.~(\ref{eq:bathcorExp}) for the calculation of the absorption spectra of Fig.~\ref{fig:Absorption}.}
\end{table}
We consider a linear chain consisting of $N=4$ identical molecules.
Taking only nearest-neighbor interactions into account the system Hamiltonian (\ref{eq:H_sys}) reads $H=\sum_{\ell=1}^N \epsilon \ket{\ell}\bra{\ell} + \sum_{N=1}^3 V (\ket{\ell}\bra{\ell+1}+ \ket{\ell+1}\bra{\ell})$. 
In the following we present results for the case $V=600\, \mathrm{cm}^{-1}$  and we present all spectra shifted by the irrelevant total energy $\epsilon$.
The spectral density of all monomers is taken to be equal and is chosen as
\begin{equation}
C(\omega)=p\Big( \frac{1}{(\omega-\Omega)^2+ \gamma^2} -\frac{1}{(\omega+\Omega)^2+ \gamma^2}\Big)
\end{equation}
with $\Omega= 500\, \mathrm{cm}^{-1}$ and $\gamma=0.01\,\Omega=5\, \mathrm{cm}^{-1}$ and $p=1.2\times 10^6$ (which corresponds to a reorganization energy of $E_{\rm r}=\frac{1}{\pi} \int_0^\infty d\omega C(\omega)/\omega \approx 500\, \mathrm{cm}^3$). 
We show calculations for $T=300\, \mathrm{K}$. 
The applied bath-correlation function Eq.~(\ref{eq:bathcorExp})  has $J=3$ terms which are provided in Table~\ref{tab:BCF_abs_params}. 
The chosen parameters are quite challenging because the spectral density represents a vibrational mode that is roughly resonant with energy differences of the electronic system ($\Omega=V$), the bath-correlation function is slowly decaying (small $\gamma$)  and the coupling to the system is large (large $p$).  
Therefore a large depth is needed to obtain converged results.
For the calculations shown we have used a depth $D=13$;
 for the triangular truncation scheme this results in a difference from the $D=12$ results which is no longer visible on our plotted resolution.  
  This requires $5\times 10^6$ auxiliary states in the triangular truncation without additional approximations.
  We will refer to this TT calculation in the following as being converged.
 
All calculations are performed using the formalism described in Ref.~\cite{ZhLiEi17_e25386_}, section 3.1. 
In particular, in the following we plot the frequency dependence of the transition strength (Eq.~(5) of Ref.~\cite{ZhLiEi17_e25386_}), which we will denote in the following simply by 'absorption'.
Recall that there is no stochasticity involved in the calculation of the absorption spectra.

\begin{figure}
\includegraphics[width=7.6cm]{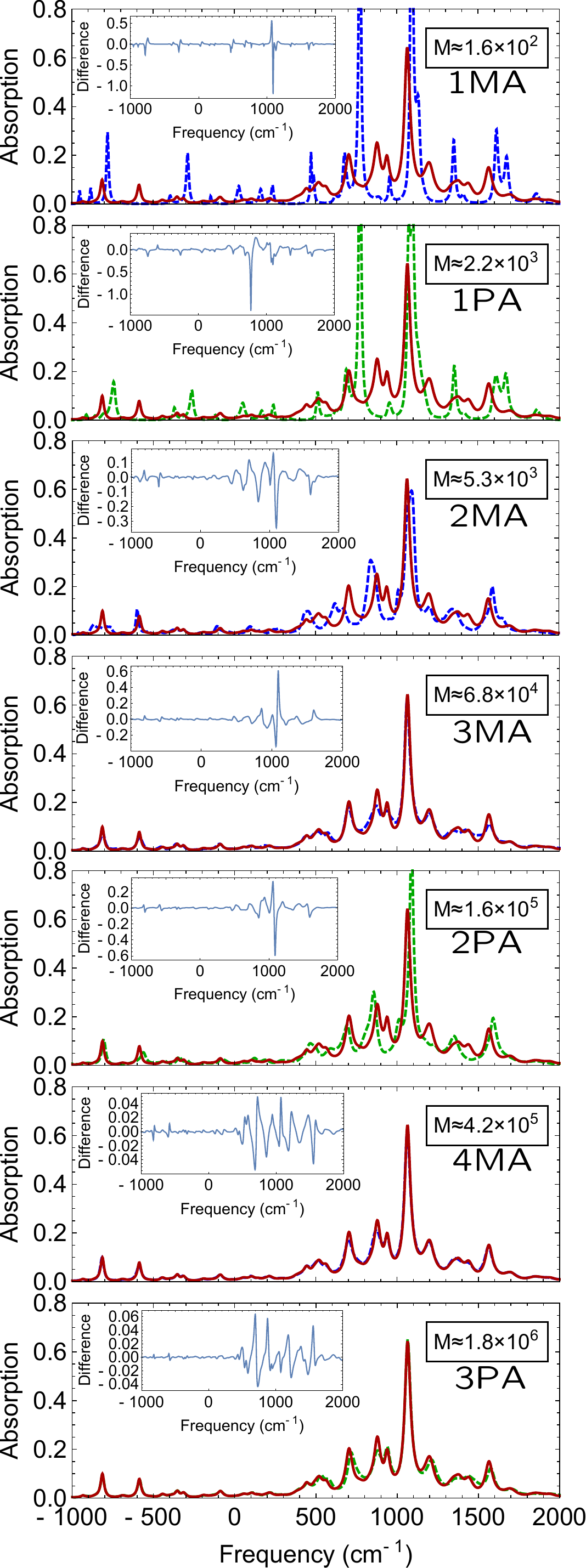}
\caption{\label{fig:Absorption}
Absorption spectra for the different approximations as indicated in the subplots, with the respective number of auxiliary equations.
In all subplots the red curve is the converged TT calculation with $D=13$.
All spectra are normalized to the same area. The zero of frequency is at the electronic transition frequency $\epsilon/\hbar$.
Further details are provided in the main text.
The inset shows the difference between the approximation used in the respective subplot and that of the subplot immediately below. For the subplot at the bottom, the difference to the red curve is shown. Note the different scales of the vertical axes of the insets.
}
\end{figure}

In Fig. \ref{fig:Absorption} we show a comparison of the different approximations (indicated in each subplot) with the converged  TT calculation (red curves).
Here one sees  that for the 1MA and the 1PA there is little agreement with the converged calculations. One should note that the respective number of auxiliary states are quite small (156 and 2236).
However, already for the 2MA with only around 5500 auxiliary equations one finds that the main features of the spectrum become visible.
Remarkably, the 3MA (which has around $70,000$ auxiliary states; approximately half as many equations as the 2PA) already has very good agreement with the converged result.
When going to the 2PA the results become slightly less accurate. 
This shows the relevance of the different kinds of states involved in the \nPA{} and \nMA{} schemes.
 For 4MA and 3PA one has reached quite high accuracy and there is only a small difference between the two spectra.  
 Note that within a certain approximation scheme (either \nMA{} or \nPA{}), the results become better with increasing order.

\subsection{Example 2: Energy transfer in the FMO complex} \label{sec:FMO}

\begin{figure*}[btpb]
\includegraphics[width=12cm]{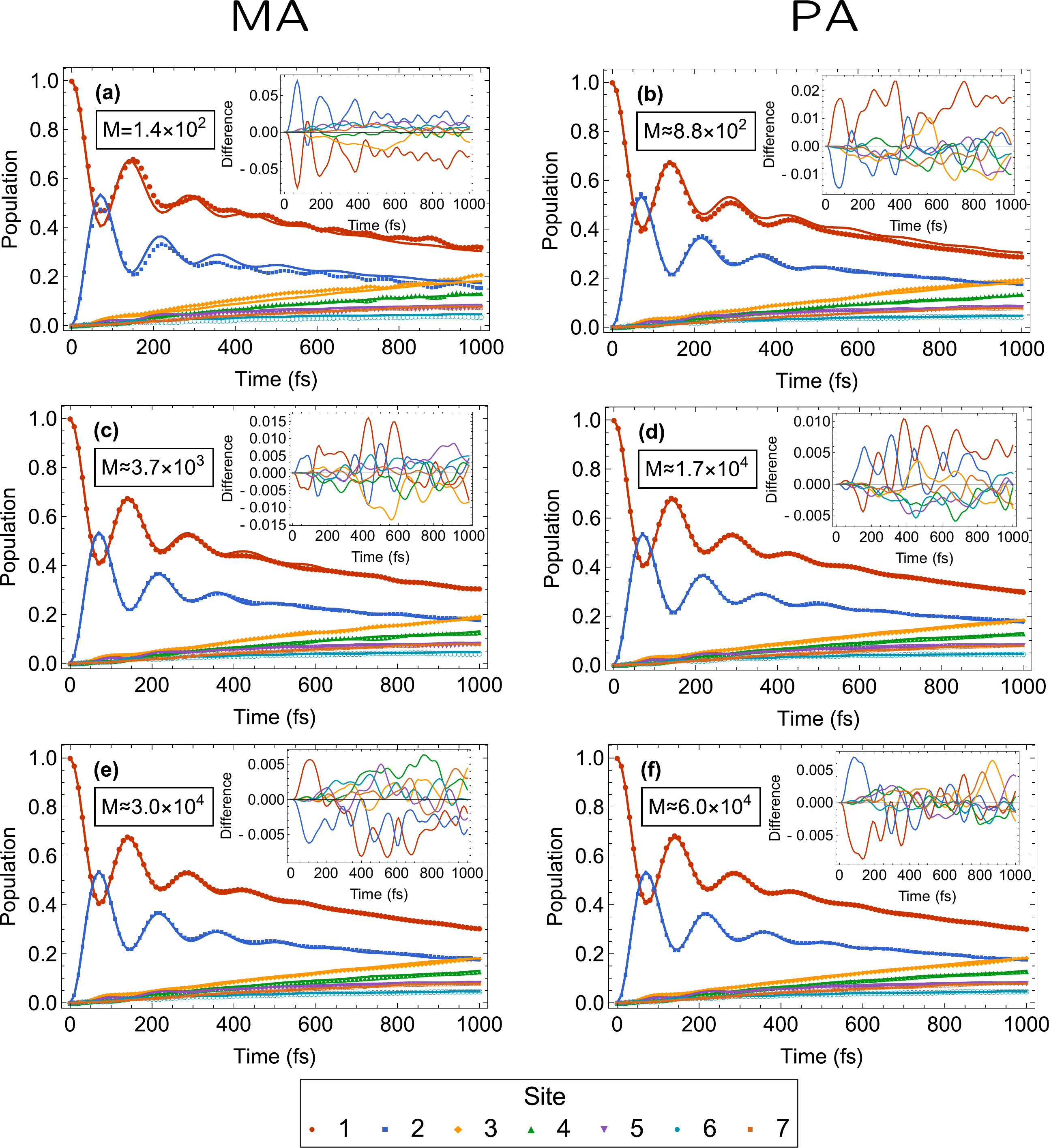}
\caption{\label{fig:depth2_nMA} Transport in the FMO complex using different truncation schemes (dotted curves). In the left/right column calculations using the \nMA{}/\nPA{} are shown (first row: 1MA, 1PA; second row 2MA, 2PA; third row: 3MA, 3PA).   
For comparison, in all panels the TT results are shown($M\approx 8.2 \times 10^4$).
In all calculations the depth is $D=4$.
The inset shows the difference between calculations with an increasing number of auxiliary states. This means that in subplot (a) we show the difference between 1PA and 1MA, in subplot (b) we show the difference between 2MA and 1PA, and so on.
In the last plot we show the difference between TT and 3PA.
In all calculations the Hamiltonian and the spectral density as specified in the main text have been used. The temperature is $300$\, K. The number of stochastic trajectories is $10000$.
Note the different scales of the vertical axes of the insets.}
\end{figure*}

\begin{table}[b]
     \centering
     \begin{tabular}{c*{2}{r}r}
         $\quad j \quad$              & $p_j\quad\quad $& $\quad\quad\Omega_j\quad $ & $\quad \quad\gamma_j\quad $ \\
         \hline
         1                & $3000-1670\,i$   & $-55$ & 52    \\
         2                & $-177-183\,i$ &$-212$    &138    \\
         3                & $4300+1150\,i$    & 55  &52         \\
         4                & $-76+700\,i$   & 212 & 138    \\
         5                &0.317    &0.0 &1615    \\
         \hline
     \end{tabular}
     \caption{Parameters for the bath-correlation function of the background spectral density in the main text, section~\ref{sec:FMO}.}\label{tabel:Backfround}
 \end{table} 

The FMO complex has been used in many theoretical studies to investigate the performance of numerical methods. 
For the calculation we choose the Hamiltonian derived in Ref.~\onlinecite{AdRe06_2778_} (given as Table S1 in Ref.~\onlinecite{IsFl09_17255_}), because most theoretical studies have used this Hamiltonian.  
For this Hamiltonian one has $N=7$.
We choose a  {\it log-normal} form for the spectral density
$
C_{\rm bg}(\omega)=\frac{\pi S \omega}{\sqrt{2\pi}\sigma} e^{-[\ln(\omega/\omega_{\rm c}]^2/2\sigma^2}.
$

Such a spectral density has been suggested to describe the broad background obtained when experimentally extracting the spectral densities of bacteriochlorophyll molecules in pigment-protein complexes~\cite{KeFeRe13_7317_}.
The numerical values for the parameters (taken from Ref.~\cite{KeFeRe13_7317_}) are $S=0.3$, $\sigma=0.7$, and $\omega_c=38\,{\rm cm}^{-1}$.

We represent the corresponding bath correlation function for the log-normal spectral density at 300 K by a sum of $J=5$ exponentials  (see Ref.~\cite{RiEi14_094101_}). The parameters are provided in table~\ref{tabel:Backfround}.

In Fig.~\ref{fig:depth2_nMA}, exemplary calculations are shown that demonstrate the convergence with increasing number of auxiliary states.
All calculations are performed using 10000 trajectories.
In the different panels the quality of the different approximation schemes is shown.
In the left column we show \nMA{} calculations and in the right column \nPA{}. From top to bottom we increase $n$, i.e., in the first row we show 1MA and 1PA, in the second row 2MA and 2PA and in the third row 3MA and 3PA.
In each panel the number of auxiliary states $M$ is provided; the corresponding curves are plotted as dotted lines.
In all panels the solid lines are a calculation performed with $D=4$ and the TT truncation scheme.
These curves serve as our reference.
We have found that there is only very little improvement by going from $D=3$ to $D=4$, we do not observe any improvement for the applied number of trajectories.
The insets show the difference between calculations with increasing numbers of auxiliary states. This means that in subplot (a) we show the difference between 1PA and 1MA, in subplot (b) we show the difference between 2MA and 1PA, etc.
In the last plot we show the difference between TT and 3PA.

One sees that the 1MA with only 140 auxiliary states already gives reasonable agreement with the 'subsequent' approximation 1PA.
However, differences are clearly visible. For example the first oscillation is not reproduced well, which can be seen in the maximal error of 0.06 in the inset.
The next lowest number of auxiliary states is from the 1PA (875).
Here one observes considerable improvement for short times ($t<200$ fs).  At later times there is a maximal deviation from the 2MA approximation of about 0.025.
For the 2MA ($\sim$ 3700 auxiliary states) we already see very good agreement up to 400 fs (maximal error 0.01).
For longer times the error is also quite small.
Further minor improvement occurs when going to the 2PA ($\sim$ 17000 equations).
Upon increasing the number of auxiliary states further, we do not find additional clear improvement, which is related to the accuracy provided by the number of trajectories.

\section{Conclusions} \label{sec:conc}
In the present paper we have discussed and compared different schemes to truncate the stochastic hierarchy of pure states, HOPS, in the context of energy transfer and absorption of molecular aggregates.
In particular we have considered the applicability of the {\it n-particle approximation} \nPA{} (section \ref{sec:nPA}) and we have introduced a novel scheme denoted by {\it n-mode approximation}, \nMA{} (section \ref{sec:nMA}).
We have discussed the number of auxiliary equations resulting from the different schemes as a function of the relevant system and bath parameters (number of molecules and number of exponentials ('modes') that are necessary to describe the local bath-correlation functions of each molecule).

We found that both the \nPA{} and the \nMA{} provide quite accurate results for a small number of auxiliary equations.
Our results indicate that the \nMA{} performs slightly better than the \nPA, however, to make a definite statement more studies are necessary.
A particular emphasis was on the question of finding a practical scheme that allows the performance of convergence checks.
For suitable convergence checks we require that one can increase the number of auxiliary equations in steps that are small enough that the computational effort remains reasonable, but the steps should be large enough that one sees improvement from one step to the next.
We think that a combination of the \nPA{} and the \nMA{} fulfills this requirement
 and is well suited to performing calculations for large molecular aggregates.
In the present work we have shown such convergence checks for a fixed depth of a triangular truncation scheme.
In practice one would also successively increase this depth.

In the present work we have demonstrated our results for the form of HOPS derived in Ref.~\onlinecite{SuEiSt14_150403_}, without using the 'terminator' suggested in that work, but simply setting the last term of Eq.~(\ref{eq:hops}) equal to zero when  truncating the hierarchy.
We believe that the basic features that we have seen in our present investigations will also hold for a more advanced terminator, or when using slightly different ways to treat the 'noise'.
In the present work, following Ref.~\onlinecite{SuEiSt14_150403_},  both quantum and classical noise are treated on the same footing.
Our findings should also apply for variants of HOPS where the zero-temperature bath-correlation function is used for the hierarchy and 
temperature is included via a stochastic Hermitian contribution to the system Hamiltonian
\cite{HaSt17_5834_}.

Since the HEOM method is closely related to HOPS \cite{SuStEi15_1408_}, we suspect that the \nMA{} will perform similarly well for HEOM and that our proposed scheme for convergence checks is also suitable for HEOM.
We would like to note that for small problems HEOM might be preferable over HOPS. 
However, for large problems HOPS will result in a much smaller size of the problem to solve numerically.

We have considered energy transfer and absorption of molecular aggregates with local molecular environments.
Our results will be directly applicable to similar situations like the transfer of a single electron in organic crystals. 
We also believe that the approach is suitable for cases with off-diagonal system-environment coupling or for couplings to common environments.

\bibliographystyle{journal_v5}

\end{document}